\documentclass[12pt]{article}
\usepackage{epsf}
\usepackage{epsfig}
\usepackage{cite}
\usepackage{amsmath}
\usepackage{amssymb}
\usepackage{wasysym}

\newcommand{\bea}{\begin{eqnarray}}
\newcommand{\eea}{\end{eqnarray}}
\newcommand{\ba}{\begin{array}}
\newcommand{\ea}{\end{array}}

\def\gappeq{\mathrel{\rlap {\raise.5ex\hbox{$>$}}
{\lower.5ex\hbox{$\sim$}}}}
\def\lappeq{\mathrel{\rlap{\raise.5ex\hbox{$<$}}
{\lower.5ex\hbox{$\sim$}}}}

\begin{document}
\topmargin -1.0cm
\oddsidemargin -0.4cm
\evensidemargin -0.4cm
\pagestyle{empty}
\begin{flushright}
IPPP/03/08, DCPT/03/16\\
UG-FT-147/03, CAFPE-17/03\\
March 2003\\
\end{flushright}
\vspace*{5mm}
\begin{center}

{\Large\bf Flavour Changing Neutral Currents in}

{\Large\bf Intersecting Brane Models}

\vspace{1.4cm}
{\sc S.A. Abel$^1$, M. Masip$^2$ and J. Santiago$^1$}\\
\vspace{.5cm}
{\it $^1$IPPP, Centre for Particle Theory}\\
{\it University of Durham}\\
{\it DH1 3LE, Durham, U.K.}\\
\vspace{.5cm}
{\it $^2$Centro Andaluz de
F\'\i sica de Part\'\i culas Elementales (CAFPE) and}\\
{\it Departamento de F\'\i sica Te\'orica y del Cosmos}\\
{\it Universidad de Granada}\\
{\it E-18071 Granada, Spain}\\

\end{center}
\vspace{1.4cm}
\begin{abstract}

Intersecting D-brane models provide an attractive
explanation of family replication in the context of string theory. We
show, however, that the localization of fermion families at
different brane intersections in the extra dimensions introduces
flavour changing neutral currents mediated by the Kaluza-Klein
excitations of the gauge fields. This is a generic feature in
these models, and it implies stringent bounds on the mass of the lightest
Kaluza-Klein modes (becoming severe when the compactification
radii are larger than the string length). We present the full string
calculation of four-fermion interactions in models with intersecting
D-branes, recovering the field theory result. This reveals other
stringy sources of flavour violation, which give bounds that are
complementary to the KK bounds (i.e. they
become severe when the compactification radii are comparable
to the string length). Taken together these bounds imply that the
string scale is larger than $M_s\gtrsim 10^2$ TeV,
implying that non-supersymmetric cases are
phenomenologically disfavoured.

\end{abstract}


\vfill

\eject

\pagestyle{empty}
\setcounter{page}{1}
\setcounter{footnote}{0}
\pagestyle{plain}


\section{Introduction}

String theory is, to date, the only known candidate for a
consistent description of gauge and gravitational
interactions. However, there is a large multiplicity of
possible vacua and, in general, explicit calculations are
difficult to perform. Consequently it is unlikely that we will
be able to identify the ``correct'' string vacuum from first principles,
in order to confront string theory directly with
observed physics. As an alternative, we can scan different
models and single out the cases that reproduce at low energies
the main features of the Standard Model (SM) such as non-abelian gauge
interactions, three families of chiral fermions, hierarchical
Yukawa interactions, and so on.
In particular, Calabi-Yau or orbifold compactifications
of the heterotic string have been extensively studied.
The realisation of D-branes as dynamical
objects of string theory~\cite{Polchinski:1995mt} has opened new
avenues in the search for phenomenologically viable
string models. The elusive property of chirality in the SM can for example
be accomplished in open string models either by locating D-branes at
singularities~\cite{branesatsingularities} or by allowing D-branes
to intersect at non-trivial angles~\cite{Berkooz:1996km}. The
latter possibility, which is the subject of this paper, has proven
very promising and a great deal of effort has been devoted
to the study of viable models and their
phenomenology~\cite{Aldazabal:2000dg,Blumenhagen:2000wh,Ibanez:2001nd,
Kokorelis:2002iz,Cremades:2002dh,Cvetic:2001tj,Cvetic:2002qa,Cremades:2002te,Blumenhagen:2002wn,
cosmology,Ghilencea:2002da,Cremades:2003qj,Lust:2003ky,Blumenhagen:2003vr}.

In this paper we want to address an important phenomenological
issue in models of intersecting D-branes that has not yet been
considered. Branes wrapping compact cycles intersecting at
non-trivial angles give rise to several copies of chiral fermions
living at the intersections. This multiplicity is generally
thought to be an attractive feature of these models, since it
leads to a nice explanation for family replication. However, the
different intersection points are localized at separate points in
the target space, leading to fermion non-universal couplings to
the gauge boson Kaluza-Klein (KK) excitations thereby inducing
Flavour Changing Neutral Currents (FCNCs) in the physical basis.
(This property was first pointed out in the context of brane world
models in Ref.~\cite{Carone:1999nz}.) The appearance of FCNCs,
which is a quite generic feature of models with intersecting
D-branes, is particularly relevant here because most of the
favoured models have to date been non-supersymmetric.
(Models with intersecting D-branes may have a non-minimal Higgs
structure which can also lead to the presence of FCNC. This issue
is however very model dependent and can be absent in particular models.
Thus we will not pursue its study further here.)
The stringent experimental bounds on these couplings compromise the necessarily
low scale of non-supersymmetric models, so that supersymmetric
set-ups (which are unfortunately rather hard to find) are greatly
preferred.

In the next Section, we review the generation of FCNCs induced by
the massive KK modes of bulk gauge bosons when fermions are
localized at displaced points in periodic extra dimensions. The
case of orbifold compactification was studied in detail in the
second reference of~\cite{Carone:1999nz} (see also
Ref.~\cite{Kaplan:2001ga}). In Section~\ref{Intersecting:Dbranes}
we describe the main features of models with intersecting D-branes
and study in a quantitative way the amount of FCNCs induced by the
KK excitations of gauge bosons. We also show how this field
theoretic calculation can be derived in a full string calculation
of flavour-violating four fermion interactions.
In Section~\ref{conclusions},
we discuss other
stringy sources of flavour violation beyond the KK one. Taken
together, the resulting bounds provide a strong constraint on the
string scale, $M_s\gtrsim 10^2$ TeV.

\section{Flavour changing neutral currents from gauge boson
  Kaluza-Klein modes\label{Extra:dimensions}}

To illustrate how FCNCs appear in these models let us consider
a $\mathrm{U}(1)$
gauge field
(the generalisation to non-abelian groups is straight-forward and
does not modify qualitatively the main results regarding FCNC)
living in a 5 dimensional space, where the extra
dimension $y$ is compact and of length $L$ ({\it i.e.}, the field
lives in the world-volume of a D4-brane wrapping a 1-cycle on a
torus). Let us also suppose that fermions are four-dimensional
fields $f_i$, with the different families localized at different
points in the extra dimension $y=y_i$. The relevant part of the
Lagrangian can be written as (a sum over $i$ understood)
\[
{\cal L}_5=-\frac{1}{4} F^{MN}F_{MN} +\mathrm{i}\;\bar{f}_i \gamma^\mu
D_\mu f_i\; \delta(y-y_i)\;,
\]
where $F_{MN}=\partial_M A_N-\partial_N A_M$, $D_M=\partial_M + \mathrm{i}
g_5 A_M$ and $M,N$ run over all space-time dimensions while
$\mu,\nu=0,1,2,3$. The dependence on the extra dimension of the
gauge fields $A_\mu$ can be expanded as
\[
A_\mu(x,y)=\frac{1}{\sqrt{L}}A^{(0)}_\mu(x) +\sqrt{\frac{2}{L}}
\sum^{\infty}_{n=1} \left( \cos \frac{2\pi n y}{L} A^{(n)}_\mu(x)
+ \sin \frac{2\pi n y}{ L} A'^{(n)}_\mu(x)\right)\;.
\]
Integrating the action over $y$ we obtain a 4-dimensional theory
(in the unitary gauge\footnote{We do not write the KK expansion of
the component of the gauge boson along the extra dimension since
it is not relevant for our calculation. The massive modes are the
Goldstone bosons associated to the five-dimensional gauge
invariance broken by the compactification (they decouple in the
unitary gauge), whereas the zero mode couples universally and does
not generate any FCNCs.}):
\bea \mathcal{L} &=& -\frac{1}{ 4}
\left[ F^{(0)2}_{\mu\nu} + \sum^{\infty}_{n=1} \left(
F^{(n)2}_{\mu\nu} + F'^{(n)2}_{\mu\nu}\right) \right] +\frac{1}{
2} \sum^{\infty}_{n=1} \left( \frac{2\pi n}{ L} \right)^2
\left( A^{(n)2}_\mu + A'^{(n)2}_\mu \right)\nonumber \\
&&+\mathrm{i}\;\bar{f}_i \gamma^\mu \left[ \partial_\mu +\mathrm{i}
g A^{(0)}_\mu
+ \mathrm{i} g \sqrt{2} \sum^{\infty}_{n=1} \left( A^{(n)}_\mu \cos
\frac{2\pi n y_i}{ L} + A'^{(n)}_\mu \sin \frac{2\pi n y_i}{
L}\right) \right] f_i\;,
\eea
where we have defined $g\equiv
g_5/\sqrt{L}$. Notice that there are two KK excitations,
$A^{(n)}_\mu$ and $A'^{(n)}_\mu$, at each mass level $M_n=2\pi n/
L$ and that the couplings of the different fermions to these
excitations depend on the position of the fermion in the extra
dimension: they are $g^n_i=\sqrt{2}g\cos (M_n y_i)$ and
$g'^n_i=\sqrt{2}g\sin (M_n y_i)$, respectively. These
flavour-dependent couplings will generate FCNCs in the basis of
mass eigenstates \cite{Carone:1999nz}. The coupling of the two KK
modes at the $n-$th level to the fermions can be written in terms
of currents as
\[
\mathcal{L}_n=A^{(n)}_\mu J^{(n)\mu}+A^{\prime(n)}_\mu
J^{\prime(n)\mu}.
\]
Under a unitary transformation $f_i=U_{ia} f_a$ from current
eigenstates $(f_i)$ to mass eigenstates $(f_a)$, $J^{(n)\mu}$
becomes
\[
J^{(n)\mu}=\bar{f}_a \gamma^\mu U^\dagger_{ai} g^n_i U_{ib} f_b,
\]
with a similar expression for $J^{\prime(n)\mu}$. If $g^n_i\neq
g^n_j$ the product of unitary matrices does not cancel and FCNCs
are generated. We can then integrate out the heavy KK modes and
obtain (at first order in $M_n^2$) flavour violating four-fermion
contact interactions
\begin{equation}
\mathcal{L}^{4f}=-\frac{1}{ 2} \sum_n \frac{J^{(n)\mu}
J^{(n)}_\mu+ J^{\prime(n)\mu} J^{\prime(n)}_\mu }{M_n^2}\;.
\label{contact:4f:general}
\end{equation}
Note that the four fermion amplitudes will be a sum of
contributions proportional to
\begin{align}
\sum_{n=1}^\infty \frac{g^n_i g^n_j+g'^n_i g'^n_j}{ M_n^2} =&
2 g^2 \sum_{n=1}^\infty \frac{\cos[M_n (y_i-y_j)]}{ M_n^2}\nonumber \\
=& \Big( \frac{g L}{2\pi}\Big)^2 \Big[
\mathrm{Li}_2(\mathrm{e}^{2\pi \mathrm{i} (y_i-y_j)/L})
+\mathrm{Li}_2(\mathrm{e}^{-2\pi \mathrm{i} (y_i-y_j)/L})
\Big],\label{coeff:relative:distances}
\end{align}
where $\mathrm{Li}_n(z)=\sum_{k=1}^\infty z^k/k^n$ is the
polylogarithm function and we have used the explicit form of the
KK masses in the second line. The presence of the second tower of
KK modes is necessary to preserve the global translation invariance
of the circle in such a
way that only the relative distances between fermions are
observable. The extension to $D>1$ extra dimensions can be
trivially obtained from the expressions above (except for the last
equality in Eq.~(\ref{coeff:relative:distances})), by changing the
coordinate $y\to \vec{y}\equiv (y_1,\ldots,y_D)$, the length
factor to the volume factor $\sqrt{L}\to \sqrt{V}$, the KK indices
$n\to \vec{n}=(n_1,\ldots,n_D)$, the masses $M_n\to
\vec{M}_{\vec{n}}=2\pi(n_1/L_1,\ldots,n_D/L_D)$ and by extending the
sum to one over a hemisphere on the $D-$dimensional lattice. For higher
dimensional branes the sum over KK modes diverges and requires a
UV cut-off (radiative corrections could also act as
cut-off~\cite{Masip:2000xy}).
The full string calculation, to be presented
in Section~\ref{stringtheorycalculation},
provides a natural cut-off in terms of the string scale, which we
will adopt in the field theory calculation in next Section.

To be more definite, let us calculate the $\Delta S=2$ operators
involved in flavour changing and CP violation in the Kaon system,
since they are expected to give the strongest constraint on the
compactification scale $L$~\cite{Carone:1999nz}. The relevant
fermionic currents coupling to the KK excitations of the gluon
$G^A_\mu$ are, for a general number of extra dimensions,
\begin{equation}
J^{A(\vec{n})}_\mu= (U^\dagger_{d_L})_{di} g^{\vec{n}}_{i}
(U_{d_L})_{is} \bar{d}^\alpha_{L} T^A_{\alpha\beta} \gamma_\mu
s^\beta_{L} + \mathrm{L\to R} + {\rm h.c.},
\end{equation}
where $g^{\vec{n}}_i=\sqrt{2}g_3\cos (\vec{M}_{\vec{n}}\cdot
\vec{y}_i)$, $\alpha,\beta$ are colour indices and $T^A_{\alpha\beta}$
are the fundamental representation matrices of $SU(3)$.
A similar expression holds for $J^{\prime A(n)}_\mu$.
The resulting four-fermion Lagrangian reads
\begin{align}
-\mathcal{L}^{\Delta S=2}= & \sum_{\vec{n}}{^\prime} \Big[
\frac{c^{(\vec{n})}_{LL}}{M_n^2} (\bar{d}^\alpha_{L} \gamma^\mu
s^\alpha_L) (\bar{d}^\beta_{L} \gamma_\mu s^\beta_L) +
\frac{c^{(\vec{n})}_{RR}}{M_n^2} (\bar{d}^\alpha_{R} \gamma^\mu
s^\alpha_R) (\bar{d}^\beta_{R} \gamma_\mu s^\beta_R)
\nonumber \\
& \phantom{-\sum_{n}}+ \frac{c^{(\vec{n})}_{LR}}{M_n^2}
(\bar{d}^\alpha_{L} s^\beta_R) (\bar{d}^\beta_{R} s^\alpha_L) +
\frac{\tilde{c}^{(\vec{n})}_{LR}}{M_n^2}(\bar{d}^\alpha_{L}
s^\alpha_R) (\bar{d}^\beta_{R} s^\beta_L) +{\rm h.c.} \Big] \;,
\end{align}
where prime in the sum means that it has to be extended over a
hemisphere of the $D-$dimensional lattice.
The coefficients are
\begin{equation}
c^{(\vec{n})}_{LL}=\frac{g_3^2}{3} \,\delta^{-l_s^2 M_{\vec{n}}^2} \sum_{ij}(U^\dagger_{dL})_{di}
(U_{dL})_{is} (U^\dagger_{dL})_{dj} (U_{dL})_{js}
\cos[\vec{M}_{\vec{n}}\cdot
(\vec{y}^L_i-\vec{y}^L_j)], \label{cLL}
\end{equation}
(similarly for $c^{(\vec{n})}_{RR}$ with $L\to R$) and
\begin{equation}
\tilde{c}^{(\vec{n})}_{LR}=-3c^{(\vec{n})}_{LR}=- 2 g_3^2
\,\delta^{-l_s^2 M_{\vec{n}}^2} \sum_{ij}(U^\dagger_{dL})_{di}
(U_{dL})_{is} (U^\dagger_{dR})_{dj} (U_{dR})_{js}
\cos[\vec{M}_{\vec{n}}\cdot (\vec{y}^L_i-\vec{y}^R_j)].\label{cLR}
\end{equation}
We have included a suppression of the gauge
coupling, $g^2\to g^2\, \delta^{-l_s^2 M_{\vec{n}}^2}$ with
$\delta\gtrsim 1$ and $l_s$ the string length,
acting as a cut-off at the
string scale. This type of suppression will be derived in the
string calculation in
Section~\ref{stringtheorycalculation}.
$U_{d_{L,R}}$ are the unitary matrices diagonalising
the down mass matrix
\[
(U_{d_L}^\dagger)_{ai} \mathcal{M}^d_{ij} (U_{d_R})_{jb}=m^d_a
\delta_{ab},
\]
and we have used $T^A_{\alpha\beta}T^A_{\gamma\delta}=
-\delta_{\alpha\beta}\delta_{\gamma\delta}/6+
\delta_{\alpha\delta}\delta_{\beta\gamma}/2$ and the Fierz
rearrengements (for anticommuting fields) $(\bar{a}_{L} \gamma^\mu
b_L)(\bar{c}_{L} \gamma^\mu d_L) =(\bar{a}_{L} \gamma^\mu
d_L)(\bar{c}_{L} \gamma^\mu b_L)$ and $(\bar{a}_{L} \gamma^\mu
b_L)(\bar{c}_{R} \gamma^\mu d_R) =-2(\bar{a}_{L} d_R)(\bar{c}_{R}
b_L)$. In the next Section we apply these results to the
calculation of FCNCs in explicit models with intersecting D-branes.

\section{Intersecting D-brane models\label{Intersecting:Dbranes}}

Models with branes intersecting at angles have received a great
deal of attention in recent years. In particular, it has
been shown that they allow the construction of models containing just the
SM spectrum and symmetries in the low energy effective theory. In
these models each stack of $N$ Dp-branes defines a
$(p+1)$--dimensional gauge theory with $U(N)\sim SU(N)\times U(1)$
symmetry~\footnote{Another possibility, phenomenologically very
appealing, is the appearance of orthogonal or symplectic groups in
the presence of orientifold fixed planes.}. (The extra $U(1)$s can
become massive by combining with RR-fields, giving rise to unbroken
global symmetries at the perturbative level. See
also~\cite{Ghilencea:2002da} for a recent study of the
phenomenology of these extra $U(1)$ gauge bosons.) Massless
fermions live in the four-dimensional intersections of two stacks
of branes, and transform under the bifundamental representation of
the corresponding groups. Chirality can be automatic, as it is in
the case of D6-branes wrapping factorizable 3-cycles in a
six-dimensional torus $T^2\times T^2\times T^2$, or obtained by
locating the intersections at orbifold fixed points, as is the
case with D(3+$n$)-branes on $n$-cycles in $T^{2n}\times
R^{6-2n}/Z_N$, with $n=1,2$~\cite{Aldazabal:2000dg}. Light scalars
live near the intersections with (possibly tachyonic) masses
depending on the particular values of the angles between branes,
allowing for \mbox{(quasi-)supersymmetric} configurations for
particular values of the
angles~\cite{Cvetic:2001tj,Cvetic:2002qa,Cremades:2002te}, and
also giving rise to fields with the quantum numbers of the Higgs
boson. The multiple wrapping of the branes around compact cycles
leads to a number of intersections which explains family
replication. As a consequence, in this kind of models
different families are necessarily localized at different space-time points.
This in turn
induces non-universal couplings to the KK excitations of the
gauge bosons and FCNCs through the mechanism described in the
previous section. The appearance of FCNCs is a quite generic
feature of models with intersecting branes. Bearing in mind that
similar features occur in more general Calabi-Yau
compactifications~\cite{Blumenhagen:2002wn}, we will focus here on
toroidal compactifications to make
the calculations tractable.

\subsection{Field Theory Calculation \label{fieldtheorycalculation}}

In order to quantify the amount of flavour violation present in
this class of models, we will concentrate on a particular case that
can be considered as a good starting point for a fully realistic
model of intersecting D-branes. It consists of an orientifold
compactification of Type IIA string theory on a 6-torus $T^2\times
T^2\times T^2$ with four stacks of D6-branes, called
\textit{baryonic} (a), \textit{left} (b), \textit{right} (c) and
\textit{leptonic} (d), giving rise to the gauge groups
$\mathrm{SU}(3),\mathrm{SU}(2)_L, \mathrm{U}(1)_R$ and
$\mathrm{U}(1)_\mathrm{lepton}$, respectively.
The orientifold projection is implemented
as $\Omega \mathcal{R}$, where $\Omega$ is the
world-sheet parity and $\mathcal{R}$ is a reflection
with respect to the first component of each 2-torus,
$\mathcal{R}Z_I=\bar{Z}_I$, with $Z_I=X_{2I+2}+\mathrm{i}
X_{2I+3}$, $I=1,2,3$.
The D6-branes have 4 extended space-time
plus three compact dimensions, each of which wraps a 1-cycle on each
of the three 2-tori. Let us denote by $(n^I_a,m^I_a)$ the
1-cycle that the $a$ stack of branes wraps, going $n^I_a$ times around the
real dimension and $m^I_a$ times around the imaginary dimension of
the complex $I-$th torus. Each brane $a$ is accompanied by an orientifold
image $a^\ast$ with wrapping numbers $(n^I_a,-m^I_a)$. Chiral
fermions live in the four-dimensional intersections between the
different branes, transforming in the bifundamental representation
of the corresponding groups, $(N_a,\bar{N}_b)$ for an $a,b$
intersection and $(N_a,N_b)$ for an $a,b^\ast$ intersection. Their
number depends on a purely topological property, the net
intersection number
\begin{equation}
I_{ab}=(n_a^1 m_b^1-m_a^1 n_b^1)\times (n_a^2 m_b^2-m_a^2
n_b^2)\times (n_a^3 m_b^3-m_a^3 n_b^3),
\end{equation}
with negative intersection numbers corresponding to a positive
number of opposite chirality fermions. Consistency conditions
given by RR tadpole cancellation plus the requirement of a
realistic massless spectrum impose stringent restrictions in the
possible configurations of intersecting D-branes, becoming
even stronger if supersymmetry is to be preserved.
Although some progress has been made in this direction,
either with extra exotic states in the
spectrum~\cite{Cvetic:2001tj,Cvetic:2002qa},
or with locally supersymmetric
models~\cite{Cremades:2002te}, in which quadratic
corrections only appear at two loop order, the search for fully realistic
supersymmetric models has proven an extremely difficult task. A final ingredient of
these models relevant for us is the structure of Yukawa couplings
between two fermions and the Higgs boson (which in the class of
models we are considering arises from the intersection between the
left and the right or the orientifold image of the right branes).
The main contribution to Yukawa couplings comes from worldsheet
instantons, given by the exponential of (minus) the area of the
worldsheet stretching between the three intersections involved.
Complex structure can appear in the Yukawa couplings when the $B$-field
or Wilson lines are turned on. (See Ref.~\cite{Cremades:2003qj}
for a recent calculation of Yukawa couplings in general toroidal
and certain Calabi-Yau compactifications.)

\begin{figure}[h]
\begin{center}
\epsfig{file=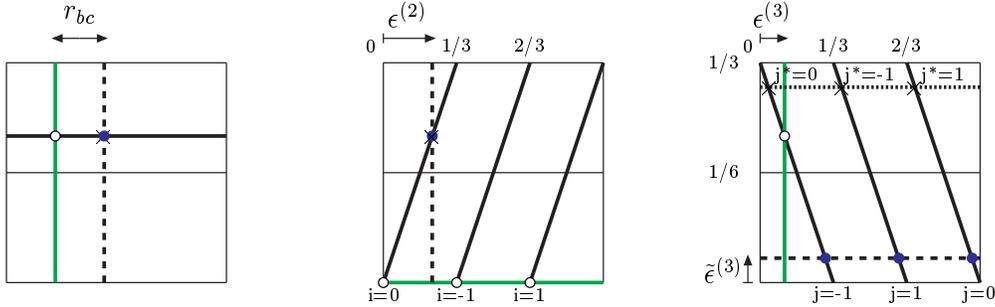,height=4cm}
\caption{Brane configuration in a model of D6-branes intersecting at angles.
The leptonic sector is not represented while the baryonic, left, right and
orientifold image of the right are respectively the dark solid, faint solid,
dashed and dotted. The intersections corresponding to the quark doublets ($i=-1,0,1$),
up type singlets ($j=-1,0,1$) and down type singlets ($j^\ast=-1,0,1$) are denoted by
an empty circle, full circle and a cross, respectively.
All distance parameters are measured in units
of $2 \pi R$ with $R$ the corresponding radius (except $\tilde{\epsilon}^{(3)}$ which is measured
in units of $6 \pi R$).\label{tori}}
\end{center}
\end{figure}

For the sake of concreteness we consider the particular model
presented recently in Ref.~\cite{Cremades:2003qj}. Although it
does not give rise to a realistic pattern of fermion masses and
mixing angles, the features that are of relevance for our discussion
are much clearer in this model than in more realistic but involved ones.
The model is
represented graphically in Fig.~\ref{tori}, where we have omitted
the leptonic sector for clarity. The relevant geometry for flavour
physics takes place in the second and third tori with no
inter-generation distances occurring in the first torus. Quark
doublets, which live at the intersections between the baryonic
(dark solid) and the left (faint solid) branes, are labelled in the
plot by $i=-1,0,1$. Up-type quark singlets ($j=-1,0,1$) live at
the intersections between the baryonic and the right (dashed)
branes while down-type singlets ($j^\ast=-1,0,1$) live at the
intersection between the baryonic and the orbifold image of the
right (dotted) brane. Two Higgs fields are localized at the
intersection (in the second and third tori) between the left and
the right and the image of the right branes respectively. For a
particular configuration, namely the ratio of the radii being
equal in the second and third tori $R^{ (2)}_2/R^{(2)}_1=R^{
(3)}_2/R^{(3)}_1$, the same $\mathcal{N}=1$ supersymmetry is
preserved at all the intersections and the model could in
principle be embedded in a bigger $\mathcal{N}=1$ globally
supersymmetric configuration. In particular, the massless
particles fill out the spectrum of the MSSM.

The lengths of the cycles that the $\mathrm{SU(3)}$ brane wraps on
the different tori are, respectively,
\begin{equation}
L_1=2\pi R^{(1)}_1,\quad L_{I}=2\pi
\sqrt{(R^{(I)}_1)^2+9(R^{(I)}_2)^2},\quad I=2,3.
\end{equation}
The relative locations of the different families (distances up to
integer multiples of $L_I$) are straight-forward to compute,
\begin{equation}
\begin{array}{|l|l|}
\hline
\mbox{    2nd torus} & \mbox{    3rd torus} \\
\hline
y^{(2)}_{iL}-y^{(2)}_{jL}=(1-\delta_{ij}) L_2/3, &
y^{(3)}_{iL}-y^{(3)}_{jL}=0,\\
y^{(2)}_{iR}-y^{(2)}_{jR}=0, &
y^{(3)}_{iR}-y^{(3)}_{jR}=(1-\delta_{ij}) L_3/3, \\
y^{(2)}_{iL}-y^{(2)}_{jR}=(i/3+\epsilon^{(2)}) L_2, &
y^{(3)}_{iL}-y^{(3)}_{jR}=(j/3+\epsilon^{(3)}-\tilde{\epsilon}^{(3)})
L_3,
\\\hline
\end{array}
\end{equation}
where the meaning of the different coefficients is explained in
Figure~\ref{tori} and in the last equation we have written the relative
separation for down-type quarks (the up-type case being the same with
the substitution $\tilde{\epsilon}^{(3)} \to
-\tilde{\epsilon}^{(3)}$).

Using the results in the previous Section we may write the
contribution to the mass difference and CP violation in the Kaon
system
\begin{equation}
\Delta m_K=\frac{\mathrm{Re} \langle K^0 | - \mathcal{L}^{\Delta
S=2} | \bar{K}^0 \rangle}{m_K},
\end{equation}
and
\begin{equation}
|\epsilon_K|=\frac{|\mathrm{Im} \langle K^0 | -
\mathcal{L}^{\Delta S=2} | \bar{K}^0 \rangle|}{2\sqrt{2} m_K
\Delta m_K}.
\end{equation}
In the vacuum insertion approximation, the relevant matrix element
can be written as (see for instance~\cite{Moroi:2000mr})
\begin{eqnarray}
\langle K^0 | - \mathcal{L}^{\Delta S=2} | \bar{K}^0 \rangle &=&
\sum_{\vec{n}}{^\prime} \Big[\frac{2}{3}
(c^{(\vec{n})}_{LL}+c^{(\vec{n})}_{RR}) +
 \Big( \frac{1}{2}\frac{m_K^2}{(m_s+m_d)^2}+\frac{1}{12} \Big)
 c^{(\vec{n})}_{LR}
\nonumber \\
&&\phantom{\sum_{\vec{n}}{^\prime}}+
 \Big( \frac{1}{6}\frac{m_K^2}{(m_s+m_d)^2}+\frac{1}{4} \Big)
\tilde{c}^{(\vec{n})}_{LR}
\Big] \frac{m_K^2 f_K^2}{M_{\vec{n}}^2},
\end{eqnarray}
where $m_{d,s}$ are the masses of the down and strange quarks,
respectively, and $f_K=160$ MeV is the Kaon decay constant.
Experimental constraints on the Kaon mass difference and epsilon
parameter now place strong bounds on the masses of the KK modes and
thus on the length of the cycles that the branes wrap in the
different tori. Writing the length of the cycles in terms of the
largest one, $\lambda_I=L_I/L_\mathrm{max}$, and requiring the
gluon KK contribution to be smaller than the experimental values,
$\Delta m_K= 3.5\times 10^{-15}$ GeV and $|\epsilon_K|=2.3\times
10^{-3}$, we can put a bound on the mass of the lightest gluon KK
mode,
\begin{equation}
M_{1} \geq 700\left(\frac{ \sqrt{ \sum_{\vec{n}}^{\prime} \frac{
|\mathrm{Re} [\frac{2}{3} (c^{(\vec{n})}_{LL}+c^{(\vec{n})}_{RR})
+ 7.2 c^{(\vec{n})}_{LR}+ 2.6 \tilde{c}^{(\vec{n})}_{LR}]|}
{(n/\lambda)^2 } }}{0.37} \right) \quad \mbox{TeV (from $\Delta
m_K$)},
\end{equation}
and
\begin{equation}
M_1 \geq 8800 \left(\frac{ \sqrt{ \sum_{\vec{n}}^{\prime} \frac{
|\mathrm{Im} [\frac{2}{3} (c^{(\vec{n})}_{LL}+c^{(\vec{n})}_{RR})
+ 7.2 c^{(\vec{n})}_{LR}+ 2.6 \tilde{c}^{(\vec{n})}_{LR}]|} {
(n/\lambda)^2 } }}{0.37} \right)  \quad \mbox{TeV (from
$\epsilon_K$)},
\end{equation}
where the different coefficients are defined in Eqs.~(\ref{cLL})
and~(\ref{cLR}) and we have denoted $(n/\lambda)^2\equiv
n_1^2/\lambda_1^2+ n_2^2/\lambda_2^2+ n_3^2/\lambda_3^2$. The
bounds have been normalised where the angles are of similar size to
those of the CKM matrix, $\epsilon^{(2)}=\epsilon^{(3)}=0.2$,
$\tilde{\epsilon}^{(3)}=0.3$, $\lambda_I=1$ (all cycles of the
same size) and to be conservative
we have taken only the contribution from the first KK level.
One could argue here that no bound is set on the
string scale at all, merely the compactification scale.
However the string scale should be reasonably
close to the compactification scale. It certainly cannot be much smaller if
non-negligible Yukawa couplings are to be generated, and it cannot
be much larger since the divergent contribution of
the KK modes is regulated by  a string scale cut-off
(as we shall see presently).
Furthermore, we will see in the next Section that when
the string length is of the order of the
compactification scale, new stringy sources of flavour violation again
banish the string scale to very high values.

Thus, the stringent bounds we have found on the mass
of the first KK mode can be translated into strong
bounds on the string scale, implying that non-supersymmetric
configurations are strongly disfavoured. It should be noted here
that the previous calculation should be taken as an estimate of
the order of magnitude of generated FCNCs. However,
despite the uncertainties, we have been
conservative in the actual calculation. Larger mixing angles or
a higher cut-off for the multi-dimensional sums can
significantly increase the induced FCNCs. For instance,
using the string inspired coupling suppression of higher KK modes
and taking the string scale to be $l_s=L_I/20$ and all the other
numbers as above we obtain a bound on the string scale
\begin{equation}
M_s\gtrsim \left\{ \begin{array}{ll} 3200 \quad \mathrm{TeV},& \mbox{ from $\Delta m_K$},\\
 40000 \quad \mathrm{TeV},& \mbox{ from $\epsilon_K$},
 \end{array} \right.
\end{equation}
which corresponds to $M_1\gtrsim 1000$ TeV
and $M_1 \gtrsim 12600$ TeV, respectively. Of course
the bounds obtained depend on many parameters and could be
smaller as well in particular models, so we should remark that
the expressions above are quite
general and can be applied to any model.

\subsection{The string theory calculation \label{stringtheorycalculation}}

In this Section we calculate the typical contribution to FCNC
processes in models where the chiral matter multiplets come from
the intersection of D-branes at angles. We will see that the field
theory result is recovered, along with a number of other features.
First we shall find a natural stringy explanation for the cut-off
which has to be added by hand in the field theory. Second we can
consider additional flavour changing processes such as \(
e^{-}e^{+}\rightarrow \mu ^{-}\tau ^{+} \) that come from the exchange
of stretched string modes. From the world-sheet point of
view these are additional instanton contributions.

Four fermion interactions have previously been considered for
orthogonal D branes in Ref.~\cite{Antoniadis:2000jv}. For theories
with branes at angles these processes are particularly important
because the sector of chiral fermions is rather independent of the
general set up, whereas the scalars are more model dependent and
may be tachyonic. The techniques for calculating 3 and 4 point
amplitudes with intersecting branes will be presented in detail
elsewhere \cite{ant}. Here we shall present the 4 point
calculation and extract the results necessary for the present
analysis, in particular to show that the generation dependence is
as described for the field theory.

String states that are stretched between branes at angles are
analogous to twisted states in the closed string, and much of the
calculation can be made using that technology
\cite{Hamidi:1986vh,Dixon:1986qv}. Indeed if two D-branes
intersect in a single complex dimension with a relative angle \(
\pi \vartheta \) at the origin, then the complex coordinate \(
Z(z) \) describing how the world sheet of an open string attached
to both branes is embedded has the mode expansion
\cite{Berkooz:1996km}
\begin{equation}
Z=\sqrt{\frac{\alpha '}{2}}\sum _{n}\frac{\alpha _{n+\vartheta
-1}}{n+\vartheta -1}z^{n+\vartheta -1}+\frac{\tilde{\alpha
}_{n-\vartheta }}{n-\vartheta }\overline{z}^{n-\vartheta }.
\end{equation}
 A similar mode expansion obtains for the fermions with the obvious
addition of \( \frac{1}{2} \) to the boundary conditions for NS
sectors. More generally the massless fermions of interest appear
in the Ramond sector with charges, \( q_{i=0..3} \) for the 4
complex transverse fermionic degrees of freedom given by one of
the following
\begin{eqnarray*}
q & = & (+\frac{1}{2},\, \, \vartheta _{1}-\frac{1}{2},\vartheta _{2}-\frac{1}{2},\vartheta _{3}-\frac{1}{2}),\\
q & = & (\pm \frac{1}{2},\pm \frac{1}{2},\, \, \vartheta _{2}-\frac{1}{2},\vartheta _{3}-\frac{1}{2}),\\
q & = & (\pm \frac{1}{2},\pm \frac{1}{2},\pm \frac{1}{2},\, \,
\vartheta _{3}-\frac{1}{2}),
\end{eqnarray*}
depending on the type of intersection. For example D6-branes
intersecting in \( (T_{2})^{3} \) are of the first kind.
The GSO projection leaves only one 4D spinor, and the theory is
chiral. For special values of angles (0 for example) supersymmetry
may be restored, but generally supersymmetry is completely broken,
and the scalars can be heavy or tachyonic. Fermions for
D5-branes intersecting in \( (T_{2})^{2}\times C/Z_{N} \) are of the
second kind. Initially the GSO projection will leave only half the
space time spinor degrees of freedom leaving a non-chiral theory.
Hence a further orbifolding on the 1st complex dimension is
required to get a chiral theory. Finally D4-branes
intersecting in \( T_{2}\times C_{2}/Z_{N} \) correspond to the
last choice. Again the GSO projection leaves 4 states which need
to be further projected out by orbifolding in the \( C_{2} \)
dimensions. The particular orbifoldings do not effect the Ramond
charges above so the quantum part of the amplitude will be
unaffected by it. The classical part of the amplitude depends
purely on the world sheet areas. However as the orbifolding is
orthogonal to the space in which the branes are wrapping, it
cannot affect the classical part either. The only effect of the
orbifolding is therefore in projecting out the chiralitites above.

The four fermion scattering amplitude is given by a disk diagram
with 4 vertex operators \( V^{(a)} \) on the boundary. The diagram
is then mapped to the upper half plane with vertices on the real
axis. The positions of the vertices are fixed by \( SL(2,R) \)
invariance to \( 0,x,1,\infty  \) (where \( x \) is real) as usual
so that the 4 point ordered amplitude is
\begin{equation}
(2\pi )^{4}\delta
^{4}(\sum _{a}k_{a})\, A(1,2,3,4)=\frac{-\mathrm{i}}{g_{s}l_{s}^{4}}\int
^{1}_{0}dx\langle
V^{(1)}(0,k_{1})V^{(2)}(x,k_{2})V^{(3)}(1,k_{3})V^{(4)}(\infty
,k_{4})\rangle .
\end{equation}
To get the total amplitude we have to sum over
all possible orderings
\begin{eqnarray}
A_{total}(1,2,3,4)&=&A(1,2,3,4)+A(1,3,2,4)+A(1,2,4,3) \nonumber \\
& + &
A(4,3,2,1)+A(4,2,3,1)+A(4,3,1,2).
\end{eqnarray}
The vertex operators for the fermions are of the form
\begin{equation}
V^{(a)}(x_{a},k_{a})=\mathrm{const}\, \lambda ^{a}\, u_{\alpha
}S^{\alpha }\prod _{I}\sigma ^{(I)}\, e^{-\phi
/2}e^{\mathrm{i}k_{a}\cdot Z}(x_{a}),
\end{equation}
where \( u_{\alpha } \) is
the space time spinor polarization and \( S^{\alpha } \) is the
so called spin-twist operator of the form
\begin{equation}
S^{\alpha }=\prod _{I}:\exp
(\mathrm{i}q_{I}^\alpha H_{I}):,
\end{equation}
with conformal dimension
\begin{equation}
h=\sum
_{I}\frac{q_{I}^{2}}{2}.
\end{equation}
\( \sigma ^{(I)} \) is the \( \vartheta
\) twist operator acting on the \( I-\)th complex dimension, with
conformal dimension
\begin{equation}
 h_{I}=\frac{1}{2}\vartheta _{I}(1-\vartheta
_{I}).
\end{equation}
 The calculation of the 4-point function of the bosonic twist operators
is now analogous to the closed string case \cite{Dixon:1986qv},
and the quantum part follows through with only minor
modifications.
For simplicity, consider branes
intersecting at an angle \( \pi \vartheta  \) in a sub 2-torus of
the compact space.
We find a contribution from the \( \vartheta  \)
twisted bosons of
\begin{equation}
\langle \sigma _{+}(x_{\infty })\sigma
_{-}(1)\sigma _{+}(x)\sigma _{-}(0)\rangle =\mathrm{const}\,
\frac{(x_{\infty }x(1-x))^{-\vartheta (1-\vartheta
)}}{[F(1-x)F(x)]^{1/2}},
\end{equation}
where \( F(x) \) is the hypergeometric
function
\begin{equation}
F(x)=F(\vartheta ,1-\vartheta ;1;x)=\frac{1}{\pi }\sin
(\vartheta \pi )\int ^{1}_{0}dy\, y^{-\vartheta
}(1-y)^{-(1-\vartheta )}(1-xy)^{-\vartheta }.
\end{equation}
When we collect all
the contributions together the dependence on \( \vartheta \)
cancels between the bosonic twist fields and the spin-twist fields
(the same cancellation in conformal dimension that guarantees
massless states in the Ramond sector) giving
\begin{eqnarray}
A(1,2,3,4) & = & -g_{s}\alpha^\prime
\,\mathrm{Tr}(\lambda ^{1}\lambda ^{2}\lambda ^{3}\lambda ^{4}+
\lambda ^{4}\lambda ^{3}\lambda ^{2}\lambda ^{1})
 \left[ \overline{u}^{(1)}\gamma _{\mu }u^{(2)}
 \overline{u}^{(4)}\gamma ^{\mu }u^{(3)}\right]
\nonumber \\
& \times&
\int ^{1}_{0}dx\, x^{-1-\alpha 's}(1-x)^{-1-\alpha 't}\frac{1}{[F(1-x)F(x)]^{1/2}}
 \sum e^{-S_{cl}(x)}, \label{string:amplitude}
\end{eqnarray}
where \( s=-(k_{1}+k_{2})^{2} \), \( t=-(k_{2}+k_{3})^{2} \), \(
u=-(k_{1}+k_{3})^{2} \) are the usual Mandlestam variables.

\begin{figure}[t]
{\centering
\resizebox*{0.7\textwidth}{0.35\textheight}{\includegraphics{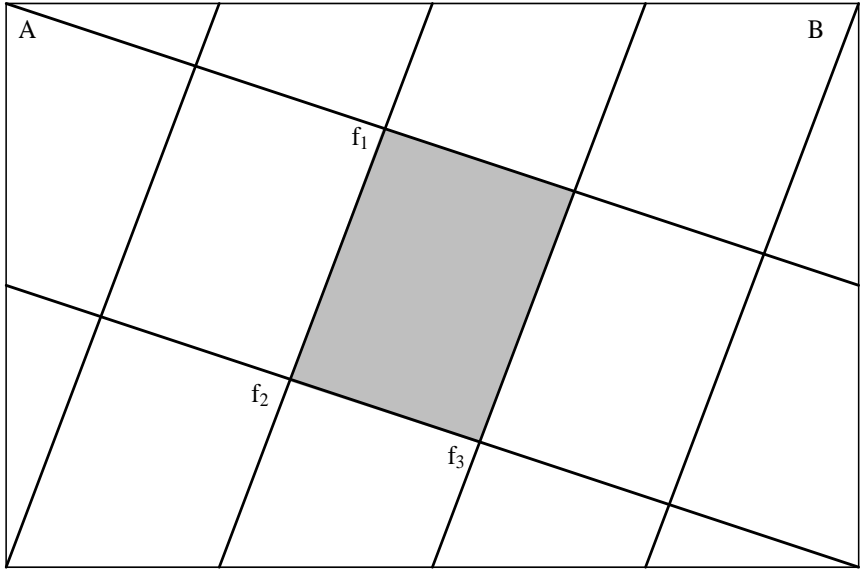}}
\par}
\caption{\label{generic}The generic 4 fermion diagram with branes
intersecting in a 2-torus.}
\end{figure}

\begin{figure}[t]
{\centering
\resizebox*{0.7\textwidth}{0.35\textheight}{\includegraphics{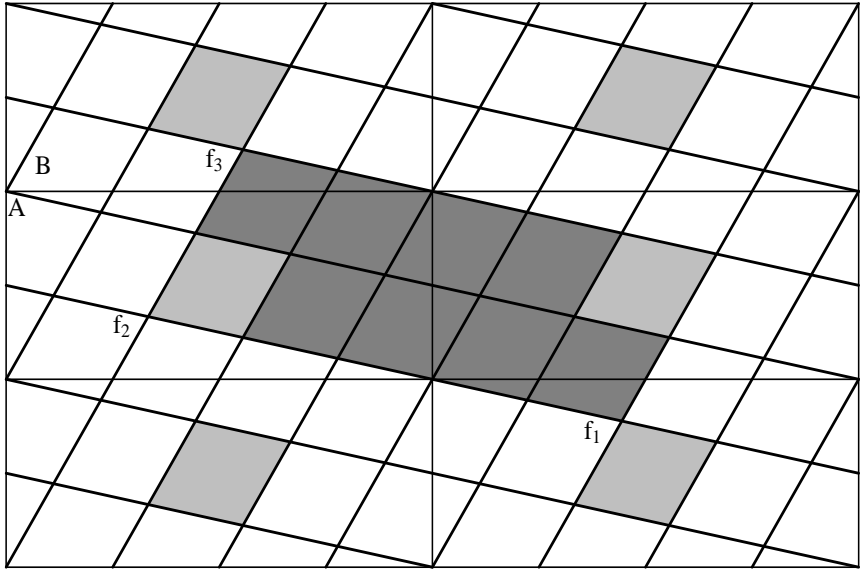}}
\par}
\caption{\label{subleading}Subleading contribution to the 4 fermion diagram with branes
intersecting in a 2-torus.}
\end{figure}

The important factor in determining the coupling is then the
instanton contribution due to the classical action which is
discussed in more detail in Ref.\cite{ant}.
Consider a generic open string four point diagram,
as shown in Fig.~\ref{generic}. In this case the classical action
turns out to be \cite{ant}
\begin{equation}
S_{cl}=\frac{\sin \vartheta \pi
}{4\pi \alpha '\tau }\left( |v'_{A}|^{2}+\tau
^{2}|v'_{B}|^{2}\right),
\end{equation}
where
\begin{equation}
v'_{A,B}=\Delta f_{A,B}+nL_{A,B},
\end{equation}
and we have defined \( \tau
(x)=\frac{F(1-x)}{F(x)} \). (For \( Z_{2} \) twists, i.e.
intersections at right-angles, this would be the modular parameter
of a {}``fake'' annulus but it has no such interpretation for more
general intersections.)
Here  $\Delta f_{A,B}$ are the displacements between consecutive
vertices along the $A$ and $B$ branes respectively,
\( n\in Z \) and \( L_{A,B} \) are the vectors in the two
torus describing the wrapped D-branes. The leading contribution in this case
comes from strings stretched
between \( f_{1} \) and \( f_{2} \) propagating along the \( A
\)-brane (in the \( f_{3}-f_{2} \) direction) as shown in Fig.~\ref{generic}
for which we choose \( \Delta
f_{A}=f_{3}-f_{2} \) and \( \Delta f_{B}=f_{1}-f_{2} \).
There is an additional contribution to the amplitude, shown in Fig.~\ref{subleading}
where \( \Delta
f_{A}=f_{1}-f_{2} \) {\em along the $A$ brane} and \( \Delta f_{B}=f_{2}-f_{3} \) {\em along
the $B$ brane}. This
corresponds to diagrams where a string stretched between \(
f_{1} \) and \( f_{2} \) along the $A$ brane propagates in the \(
f_{1}-f_{2} \) direction, \textit{i.e.} along the $B$ brane, and
because of the much larger world sheet area
is a subleading contribution for the example shown.

For diagrams where \( \Delta f_{A} \) and \( \Delta f_{B}
\) are both non-zero, we expect a world-sheet instanton
suppression factor that goes like the world-sheet area. To get
this we can use a saddle point approximation for the \( x \)
integral in \( A(1,2,3,4) \) with \( \tau
(x_{s})=|v'_{A}|/|v'_{B}| \). The accuracy of this approximation
is a function of the width of the saddle, given by \( \sqrt{4\pi
\alpha '/R^2_{c}}\sim \frac{l_{s}}{R_{c}} \), where \( R_{c} \) is
the compactification scale (\( R_{c}\sim R_{A},R_{B} \)). As
expected the approximation breaks down when the size of the world
sheet is comparable to the D-brane thickness.
Substituting back into the action we find
\begin{equation}
A(1,2,3,4)\sim \sum
_{\Lambda _{A},\Lambda _{B}}\, \mathrm{fermion-factors}\times
\left( \frac{4\pi \alpha '}{R_{c}}\right) ^{2}\exp \left(
-\frac{1}{2\pi \alpha '}\sin \vartheta \pi \,
|v'_{A}||v'_{B}|\right) . \label{string:instanton:4f:Eq}
\end{equation}

 We find the expected suppression from the mass of the intermediate
stretched string state, times by an instanton suppression given by
the area of the world sheet. These expressions will be useful in
considering more exotic flavour changing processes such as \(
ee\rightarrow \tau \mu  \).

For the moment however, we are interested in processes that do not
explicitly violate flavour, such as \( \mu ^{+}\mu ^{-}\rightarrow
e^{+}e^{-} \). For such processes the intersection separation in
the leading term \( v'_{B}=\Delta f_{ee} \) for one pair of twist
operators is zero, and so this term cannot be treated as above.
Consider the summation over \( \Lambda _{A} \) in \( v'_{A} \) for
the other pair, whose separation is \( v'_{A}=\Delta f_{e\mu
}+nL_{A} \). Poisson resumming we find
\begin{equation}
\sum e^{-S_{cl}}=\sum _{p_{A}\in \Lambda _{A}^{*}}\sqrt{\frac{4\pi
^{2}\alpha '\tau }{L^{2}_{A}\sin \vartheta \pi }}\, \exp \left[
-\frac{4\pi ^{3}\alpha '\tau }{\sin \vartheta \pi
}p_{A}^{2}\right] \, \exp \left[ 2\pi i\Delta f_{e\mu
}\cdot p_{A}\right] +\mathrm{subleading},
\end{equation}
where \(
p_{A}\in \Lambda ^{*}_{A} \)is summed over the dual lattice
\begin{equation}
p_{A}=\frac{n_{A}}{|L_{A}|^{2}}L_{A}.
\end{equation}
This expression describes
the leading exchange of gauge bosons plus their KK modes along the
\( A \) brane. (The subleading terms (those with \(
v'_{B}=n_{B}L_{B} \) with integer \( n_{B}\neq 0 \)) can still be
treated using the saddle point approximation above.) To obtain the
field theory result, we take the limit of coincident vertices, \(
x\rightarrow 0 \) or \( x\rightarrow 1 \). For example the former
contribution gives \( s \)-channel exchanges and can be evaluated
using the asymptotics
\begin{equation}
F(x)\sim 1\, ,\, \, \, \, \, \tau \sim
F(1-x)\sim \frac{1}{\pi }\sin \vartheta \pi \, \ln \frac{\delta
}{x},
\end{equation}
where \( \delta  \) is given by the digamma function \(
\psi (z)=\Gamma '(z)/\Gamma (z) \)
\begin{equation}
\delta =\exp (2\psi (1)-\psi (\vartheta )-\psi (1-\vartheta )).
\end{equation}
We find \begin{eqnarray}
A(1,2,3,4) & = & g_{s}\left[ \overline{u}^{(1)}\gamma _{\mu }u^{(2)}
\overline{u}^{(4)}\gamma ^{\mu }u^{(3)}\right]
\frac{\sqrt{4\pi ^{2}\alpha ' }}{L_{A}\sqrt{\sin \vartheta \pi} }
\nonumber \\
 &  & \times \, \left( \frac{1}{s}+2\sum ^{\infty }_{n=1}
 \frac{\cos \left( 2\pi \Delta y_{e\mu }p_{n}\right)
 \, \delta ^{-\alpha 'M_{n}^{2}}}{s-M_{n}^{2}}\right) ,
\end{eqnarray}
where \( M_{n}=2\pi n/ L_{A} \), \( p_{n}=nL_{A}/|L_{A}|^{2}
\) and we have denoted $\Delta y_{e\mu}=|\Delta f_{e\mu}|$ to
match the field theory calculation, thereby recovering the
one-dimensional case we derived in the
field theory approximation, provided that \( \alpha 'M_{1}^{2}\ll
1 \). That is, the brane separation should again be larger than
the brane thickness. Note that we have been a little sloppy
in the notation since the indices $e,\mu$ do not correspond
to flavour but to current eigenstates. This expression is thus to
be compared with Eq.(~\ref{contact:4f:general}) \textit{before} the
unitary rotations.
The extension to higher dimensional
intersecting branes follows straightforwardly, and we now find
that the form factor \( \, \delta ^{-\alpha 'M_{n}^{2}} \)
naturally provides the UV cut-off which in the field theory had to
be added by hand. Physically the cut-off arises because the
intersection itself has thickness \( \sim \sqrt{\alpha '} \), and
thus cannot emit modes with a shorter wavelength.

\section{Phenomenological discussion and conclusions\label{conclusions}}

We have seen that a general feature of models with branes
intersecting at angles is the appearance of FCNCs mediated by the
KK excitations of the gauge bosons living in the world volume of
the D-branes. The general four fermion interactions are suppressed
by the square of the compactification scale, with coefficients
\begin{equation}
c_{\mathrm{KK}}\sim \sum_n \frac{1}{M_n^2} \sim \sum_{n}
\Big(\frac{L_c}{2\pi n}\Big)^2.
\end{equation}
The experimental bounds on flavour changing neutral processes
then impose stringent constraints on the compactification scale
or, alternatively, on the mass of the lightest gauge boson KK
excitation
\begin{equation}
M_1\gtrsim 700-8800 \quad \mathrm{TeV}.
\end{equation}
This is, however, a conservative bound that arises from a purely
field theoretic calculation. Furthermore KK sums require some
regularisation for more than one extra dimension that, at this
level, has to be put by hand. We obtained a better
understanding of the situation by performing a full string
calculation of the relevant four fermion interactions. In
addition to being an interesting calculation to perform,
the string calculation provides us with a
natural regularisation of the KK sums and gives us confidence that
our physical intuition is correct. An explicit exponential
suppression of the coupling in terms of mass of the corresponding
KK mode and the string scale is found, leading to the following
regularised form of the coefficient of the four fermion
interaction
\begin{equation}
c_{\mathrm{KK}}^{\mathrm{reg}}\sim \sum_n \Big(\frac{L_c}{2 \pi
n}\Big)^2 \delta^{-(2\pi n l_s/L_c)^2},
\end{equation}
with $\delta$ a number of order one and $l_s$ the string scale.
This dramatically increases the amount of flavour violation (and thus
the bound on the compactification scale) in the limit $l_s \ll
L_c$. In the case $l_s=L_c/20$ for instance we find,
\begin{equation}
M_s \gtrsim 3200-40000 \quad \mathrm{TeV},
\end{equation}
corresponding to the mass of the first gauge boson KK mode being
\begin{equation}
M_1 \gtrsim 1000-12600 \quad \mathrm{TeV}.
\end{equation}
On the other hand, if the compactification length is of the same
order or even smaller than the string length, then only the KK modes in
the first level need to be considered, and
the bounds from this source can be much milder.

\begin{figure}[h]
\begin{center}
\epsfig{file=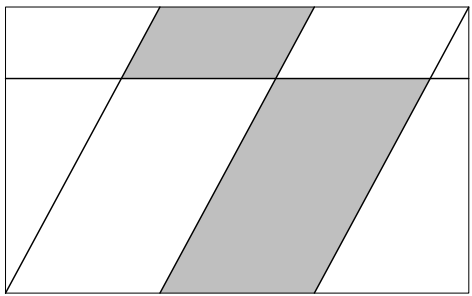,height=4cm}
\caption{Instanton contribution to flavour violating processes such as
$\tau \to ee \mu$. The string contribution is proportional to minus
the exponential of the shaded area.
\label{string:instanton:4f:fig}}
\end{center}
\end{figure}

\begin{figure}[h]
\begin{center}
\epsfig{file=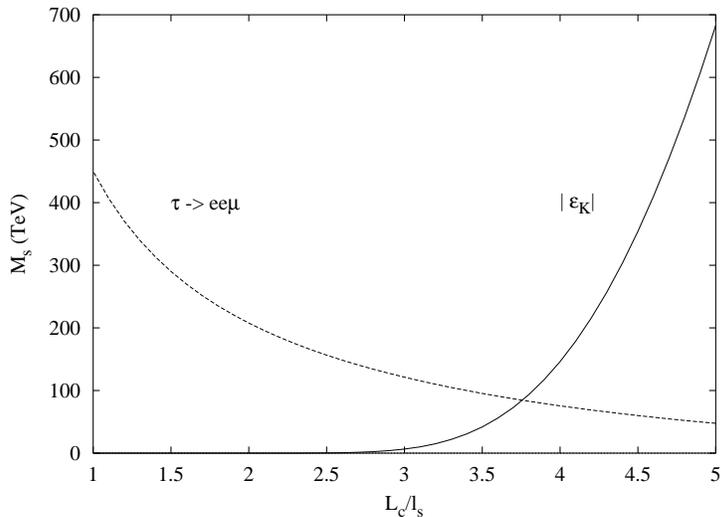,height=7cm}
\caption{Lower bound on the string scale as a function of the
  compactification length (in terms of the string length) coming from
  the contribution of gluon KK modes (solid line) and from instanton
  contributions (dashed line). \label{globalbounds}}
\end{center}
\end{figure}

The string calculation does however give us another source of
flavour violation in these models with a different dependence on
the compactification and string scales. These correspond to a process
with the
string stretched along the parallelogram formed by two sets of
intersecting D-branes, as shown in Fig.~\ref{string:instanton:4f:fig}. This can mediate
processes such as $\tau \to ee \bar{\mu}$. From the
worldsheet point of view these transitions correspond to instanton
contributions that are proportional to the area of the
parallelogram involved, see Eq.~(\ref{string:instanton:4f:Eq}). Taking all the cycles
to be of similar size, the corresponding contact interaction mediating rare
tau decays has a coefficient
\begin{equation}
c_{\mathrm{inst}}\sim \frac{8 M_Z^2 G_F}{\sqrt{2}} \frac{1}{M_s^2}
\Big(\frac{8 \pi^2 l_s}{L_c}\Big)^2 \mathrm{e}^{-A/(2\pi l_s^2)}.
\end{equation}
Using the latest Belle results on rare tau decays~\cite{Yusa:2002ff}
\begin{equation}
\mathrm{Br}(\tau\to ee\mu)\leq 3.4 \times 10^{-7},
\end{equation}
and considering small mixing angles so that the flavour eigenstates
are close to the current eigenstates we find the bounds shown in
Fig.~\ref{globalbounds}.

As Figure~\ref{globalbounds} shows, in models with intersecting D-branes,
there are stringent bounds on the string scale from
FCNC processes mediated by gauge
boson KK modes when $l_s \ll L_c$, and by stretched string states
when $l_s \gtrsim L_c$.
These bounds
seriously compromise non-supersymmetric models
which, in order to avoid a hierarchy, need a low string scale.
The bounds also imply that other phenomenological signals of
these models, such as the presence of new $U(1)$ gauge
bosons arising from the gauge reduction $U(N)\sim SU(N)\times
U(1)$ (see Ref.~\cite{Ghilencea:2002da}) are out of reach of
present or near future experiments. (Note however that in some
cases these extra $U(1)$ gauge bosons can also have family
non-universal gauge couplings~\cite{Cvetic:2002qa}, inducing FCNCs
that could become relevant. See for example Ref.~\cite{Langacker:2000ju}.)

To summarise, we have discussed how FCNCs arise in models with
intersecting D-branes, as a result of the
family non universal couplings of the KK modes of the gauge
bosons living in the world volume of the branes. We have also
presented the equivalent full string calculation, finding
good agreement with the field theoretic result.
The string calculation gives a natural regularisation of the
otherwise divergent KK sums. We also highlighted another source
of flavour violation in these models which becomes important in the
region of the parameter space (when the compactification scale is
comparable to the string scale) where the KK contribution is
subleading. A more detailed study of flavour physics in models
with intersecting D-branes is currently in progress and will be
presented elsewhere, but these initial results seem to indicate that
non-supersymmetric models are strongly disfavoured for
phenomenological reasons.

\section*{Acknowledgements}
We would like to thank Angel Uranga for
useful discussions.
We thank I. Klebanov
for pointing out an error in Eq~(\ref{string:amplitude}) in an earlier version.
This work was partially funded by PPARC and by
Opportunity Grant PPA/T/S/1998/00833.
Financial support by CICYT, Junta de Andaluc\'\i a and the European Union
under contracts FPA2000-1558, FQM-101 and HPRN-CT-2000-00149 is also
acknowledged.



\begin{thebibliography}{99}

\bibitem{Polchinski:1995mt}
J.~Polchinski,
Phys.\ Rev.\ Lett.\  {\bf 75} (1995) 4724 [arXiv:hep-th/9510017].

\bibitem{branesatsingularities}
G.~Aldazabal, L.~E.~Ibanez, F.~Quevedo and A.~M.~Uranga,
JHEP {\bf 0008} (2000) 002 [arXiv:hep-th/0005067];
D.~Berenstein, V.~Jejjala and R.~G.~Leigh,
Phys.\ Rev.\ Lett.\  {\bf 88} (2002) 071602
[arXiv:hep-ph/0105042];
L.~F.~Alday and G.~Aldazabal,
JHEP {\bf 0205} (2002) 022 [arXiv:hep-th/0203129].


\bibitem{Berkooz:1996km}
M.~Berkooz, M.~R.~Douglas and R.~G.~Leigh,
Nucl.\ Phys.\ B {\bf 480} (1996) 265 [arXiv:hep-th/9606139].


\bibitem{Aldazabal:2000dg}
G.~Aldazabal, S.~Franco, L.~E.~Ibanez, R.~Rabadan and
A.~M.~Uranga,
J.\ Math.\ Phys.\  {\bf 42} (2001) 3103 [arXiv:hep-th/0011073];
G.~Aldazabal, S.~Franco, L.~E.~Ibanez, R.~Rabadan and
A.~M.~Uranga,
JHEP {\bf 0102} (2001) 047 [arXiv:hep-ph/0011132].


\bibitem{Blumenhagen:2000wh}
R.~Blumenhagen, L.~Goerlich, B.~Kors and D.~Lust,
JHEP {\bf 0010} (2000) 006 [arXiv:hep-th/0007024];
R.~Blumenhagen, B.~Kors and D.~Lust,
JHEP {\bf 0102} (2001) 030 [arXiv:hep-th/0012156];
R.~Blumenhagen, B.~Kors, D.~Lust and T.~Ott,
Nucl.\ Phys.\ B {\bf 616} (2001) 3 [arXiv:hep-th/0107138].


\bibitem{Ibanez:2001nd}
L.~E.~Ibanez, F.~Marchesano and R.~Rabadan,
JHEP {\bf 0111} (2001) 002 [arXiv:hep-th/0105155];
D.~Bailin, G.~V.~Kraniotis and A.~Love,
Phys.\ Lett.\ B {\bf 530} (2002) 202 [arXiv:hep-th/0108131];
G.~Honecker,
JHEP {\bf 0201} (2002) 025 [arXiv:hep-th/0201037];
D.~Cremades, L.~E.~Ibanez and F.~Marchesano,
JHEP {\bf 0207} (2002) 022 [arXiv:hep-th/0203160];
J.~R.~Ellis, P.~Kanti and D.~V.~Nanopoulos,
Nucl.\ Phys.\ B {\bf 647} (2002) 235 [arXiv:hep-th/0206087];
D.~Bailin, G.~V.~Kraniotis and A.~Love,
Phys.\ Lett.\ B {\bf 547} (2002) 43 [arXiv:hep-th/0208103];
D.~Bailin, G.~V.~Kraniotis and A.~Love,
Phys.\ Lett.\ B {\bf 553} (2003) 79 [arXiv:hep-th/0210219];
D.~Bailin, G.~V.~Kraniotis and A.~Love,
arXiv:hep-th/0212112.

\bibitem{Kokorelis:2002iz}
C.~Kokorelis,
arXiv:hep-th/0212281;
%
arXiv:hep-th/0211091;
%
arXiv:hep-th/0210200;
%
arXiv:hep-th/0210004;
%
JHEP {\bf 0211} (2002) 027 [arXiv:hep-th/0209202];
arXiv:hep-th/0207234;
%
JHEP {\bf 0208} (2002) 036 [arXiv:hep-th/0206108];
%
JHEP {\bf 0209} (2002) 029 [arXiv:hep-th/0205147];
%
JHEP {\bf 0208} (2002) 018 [arXiv:hep-th/0203187].









\bibitem{Cremades:2002dh}
D.~Cremades, L.~E.~Ibanez and F.~Marchesano,
Nucl.\ Phys.\ B {\bf 643} (2002) 93 [arXiv:hep-th/0205074].



\bibitem{Cvetic:2001tj}
R.~Blumenhagen, L.~Gorlich and B.~Kors,
JHEP {\bf 0001} (2000) 040 [arXiv:hep-th/9912204];
S.~Forste, G.~Honecker and R.~Schreyer,
Nucl.\ Phys.\ B {\bf 593} (2001) 127 [arXiv:hep-th/0008250];
M.~Cvetic, G.~Shiu and A.~M.~Uranga,
Phys.\ Rev.\ Lett.\  {\bf 87} (2001) 201801
[arXiv:hep-th/0107143];
M.~Cvetic, G.~Shiu and A.~M.~Uranga,
Nucl.\ Phys.\ B {\bf 615} (2001) 3 [arXiv:hep-th/0107166];
M.~Cvetic, P.~Langacker and G.~Shiu,
Nucl.\ Phys.\ B {\bf 642} (2002) 139 [arXiv:hep-th/0206115];
R.~Blumenhagen, L.~Gorlich and T.~Ott,
JHEP {\bf 0301} (2003) 021 [arXiv:hep-th/0211059];
M.~Cvetic, I.~Papadimitriou and G.~Shiu,
arXiv:hep-th/0212177;
G.~Honecker,
arXiv:hep-th/0303015.




\bibitem{Cvetic:2002qa}
M.~Cvetic, P.~Langacker and G.~Shiu,
Phys.\ Rev.\ D {\bf 66} (2002) 066004 [arXiv:hep-ph/0205252];



\bibitem{Cremades:2002te}
D.~Cremades, L.~E.~Ibanez and F.~Marchesano,
JHEP {\bf 0207} (2002) 009 [arXiv:hep-th/0201205].


\bibitem{Blumenhagen:2002wn}
R.~Blumenhagen, V.~Braun, B.~Kors and D.~Lust,
JHEP {\bf 0207} (2002) 026 [arXiv:hep-th/0206038];
A.~M.~Uranga,
JHEP {\bf 0212} (2002) 058 [arXiv:hep-th/0208014].


\bibitem{cosmology}
J.~Garcia-Bellido, R.~Rabadan and F.~Zamora,
JHEP {\bf 0201} (2002) 036 [arXiv:hep-th/0112147];
R.~Blumenhagen, B.~Kors, D.~Lust and T.~Ott,
Nucl.\ Phys.\ B {\bf 641} (2002) 235 [arXiv:hep-th/0202124];
M.~Gomez-Reino and I.~Zavala,
JHEP {\bf 0209} (2002) 020 [arXiv:hep-th/0207278].



\bibitem{Ghilencea:2002da}
D.~M.~Ghilencea, L.~E.~Ibanez, N.~Irges and F.~Quevedo,
JHEP {\bf 0208} (2002) 016 [arXiv:hep-ph/0205083];
D.~M.~Ghilencea,
Nucl.\ Phys.\ B {\bf 648} (2003) 215 [arXiv:hep-ph/0208205].



\bibitem{Cremades:2003qj}
D.~Cremades, L.~E.~Ibanez and F.~Marchesano,
arXiv:hep-th/0302105.

\bibitem{Lust:2003ky}
D.~Lust and S.~Stieberger,
arXiv:hep-th/0302221.


\bibitem{Blumenhagen:2003vr}
R.~Blumenhagen, D.~Lust and T.~R.~Taylor,
arXiv:hep-th/0303016;
J.~F.~Cascales and A.~M.~Uranga,
arXiv:hep-th/0303024.





\bibitem{Carone:1999nz}
C.~D.~Carone,
Phys.\ Rev.\ D {\bf 61} (2000) 015008 [arXiv:hep-ph/9907362];
A.~Delgado, A.~Pomarol and M.~Quir\'os,
JHEP {\bf 0001} (2000) 030 [arXiv:hep-ph/9911252].


\bibitem{Kaplan:2001ga}
D.~E.~Kaplan and T.~M.~Tait,
JHEP {\bf 0111} (2001) 051 [arXiv:hep-ph/0110126].



\bibitem{Masip:2000xy}
M.~Masip,
Phys.\ Rev.\ D {\bf 62} (2000) 105012
[arXiv:hep-ph/0007048].




\bibitem{Moroi:2000mr}
T.~Moroi,
JHEP {\bf 0003} (2000) 019
[arXiv:hep-ph/0002208].


\bibitem{Dixon:1986qv}
L.~J.~Dixon, D.~Friedan, E.~J.~Martinec and S.~H.~Shenker,
Nucl.\ Phys.\ B {\bf 282} (1987) 13.

\bibitem{Hamidi:1986vh}
S.~Hamidi and C.~Vafa,
Nucl.\ Phys.\ B {\bf 279} (1987) 465.


\bibitem{Antoniadis:2000jv}
I.~Antoniadis, K.~Benakli and A.~Laugier,
JHEP {\bf 0105} (2001) 044 [arXiv:hep-th/0011281].


\bibitem{ant}S.Abel and A.Owen, Nucl. Phys. \textbf{B} (in press) [hep-th/0303124].



\bibitem{Yusa:2002ff}
Y.~Yusa, H.~Hayashii, T.~Nagamine and A.~Yamaguchi  [BELLE Collaboration],
arXiv:hep-ex/0211017.

\bibitem{Langacker:2000ju}
P.~Langacker and M.~Plumacher,
Phys.\ Rev.\ D {\bf 62} (2000) 013006
[arXiv:hep-ph/0001204].




\end{thebibliography}
\end{document}